\documentclass[11pt,a4paper]{article}

\usepackage{graphicx}
\usepackage{color}
\usepackage{amsmath,amsfonts,amssymb}
\usepackage{physics}
\usepackage{bbold}
\usepackage{t1enc}
\usepackage{cite}

\usepackage{amsmath,amssymb}
\usepackage{amsfonts}
\usepackage{bbold}
\usepackage{latexsym} 
\usepackage{cancel}
\usepackage{graphicx, rotating}
\usepackage{epsfig}
\usepackage{color}
\usepackage[dvipsnames]{xcolor}
\usepackage{multirow}

\newcommand{\bea}{\begin{eqnarray}}
\newcommand{\eea}{\end{eqnarray}}

\DeclareMathSymbol{\shortminus}{\mathbin}{AMSa}{"39}
\DeclareMathSymbol{\shm}{\mathbin}{AMSa}{"39}

\begin{document}

\begin{center}
\begin{Large}
{\bf Quantum tomography beyond the leading order}
\end{Large}

\vspace{0.5cm}
\renewcommand*{\thefootnote}{\fnsymbol{footnote}}
\setcounter{footnote}{0}
J. A.~Aguilar-Saavedra \\[1mm]
\begin{small}
Instituto de F\'isica Te\'orica IFT-UAM/CSIC, c/Nicol\'as Cabrera 13--15, 28049 Madrid, Spain \\
\end{small}
\end{center}
\begin{abstract}
Quantum tomography, as a tool to probe foundational aspects of quantum mechanics, relies on extracting spin information from angular distributions. This is inherently a leading-order technique, ill-defined when higher-order corrections are significant. 
For those cases, we propose to treat higher-order corrections as a background, to be modeled and subtracted from data in the  same way as other backgrounds are. We illustrate this procedure for Higgs decays $H \to ZZ \to e^+ e^- \mu^+ \mu^-$, which is of high interest for upcoming qutrit entanglement tests at the Large Hadron Collider.

\end{abstract}

\section{Introduction}

Quantum tomography of particle systems produced in high-energy collisions has recently attracted great interest as a unique tool for experimentally probing foundational aspects of quantum mechanics. The reconstruction of spin quantum states enables  tests of entanglement \cite{Afik:2020onf,Fabbrichesi:2021npl,Severi:2021cnj,Afik:2022kwm,Aguilar-Saavedra:2022uye,Afik:2022dgh,Dong:2023xiw,Han:2023fci,Aguilar-Saavedra:2023lwb,Altakach:2022ywa,Barr:2021zcp,Aguilar-Saavedra:2022wam,Aguilar-Saavedra:2022mpg,Fabbri:2023ncz,Fabbrichesi:2023cev,Morales:2023gow,Aguilar-Saavedra:2023hss,Aguilar-Saavedra:2024fig,Aguilar-Saavedra:2024hwd,Altomonte:2024upf,Fabbrichesi:2024wcd,Ehataht:2023zzt,Zhang:2025mmm,Han:2025ewp,Han:2024ugl,Maltoni:2024csn,Aguilar-Saavedra:2024vpd,Aguilar-Saavedra:2024whi,Bernal:2025zqq}, contextuality \cite{Fabbrichesi:2025ifv,Fabbrichesi:2025rsg}, and quantum interference and identical-particle effects \cite{Aguilar-Saavedra:2024jkj}. Quantum tomography relies on the fact that, for systems involving short-lived particles, the spin state is encoded in the multi-dimensional angular distributions of the decay products~\cite{Kane:1991bg,Aguilar-Saavedra:2015yza,Aguilar-Saavedra:2017zkn,Rahaman:2021fcz,Ashby-Pickering:2022umy,Bernal:2023jba}. These distributions can then be translated into measurements of spin observables, thereby allowing for the reconstruction of the spin density operator. The knowledge of the spin density operator thereby allows to perform tests of quantum mechanical properties, such as entanglement between the spins of produced particles, which has already been measured for top quark pairs~\cite{ATLAS:2023fsd,CMS:2024pts,CMS:2024zkc}.

\begin{figure}[t]
\begin{center}
\begin{tabular}{cc}
\includegraphics[height=3cm]{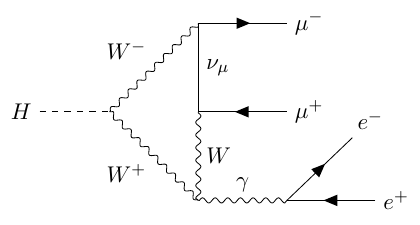}
& \includegraphics[height=4cm]{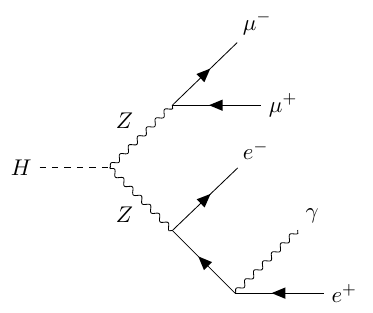} 
\end{tabular}
\caption{Examples of higher-order diagrams entering NLO (electroweak) corrections to $H \to ZZ \to e^+ e^- \mu^+ \mu^-$.}
\label{fig:diags}
\end{center}
\end{figure}

This is clearly a leading-order (LO) picture, though sufficiently accurate in many cases of interest for the near future.\footnote{Needless to say, even at LO and with distinguishable particles in the final state, a spin interpretation of angular observables is not always possible. For example, for $pp \to e^+ e^- \mu^+ \mu^-$ below the $ZZ$ threshold there is a large number of diagrams with intermediate $Z$ and $\gamma$ that significantly contribute to the amplitude. Our discussion in this paper is meant for processes for which the spin interpretation is well defined at LO.} Let us for example consider the Higgs decay $H \to ZZ \to e^+ e^- \mu^+ \mu^-$. Interpreting the angular distribution of a same-flavour pair in their centre-of-mass (c.m.) frame, in terms of spin observables of the parent $Z$ boson, is only valid at LO. At next-to-leading order (NLO) there are virtual corrections from diagrams in which the two same-flavour leptons do not result from the same $Z$ boson~\cite{Grossi:2024jae,DelGratta:2025qyp}, or even diagrams that do not have a $Z$ boson (see Fig.~\ref{fig:diags} left). Likewise, real photon emission diagrams off the $Z$ decay products invalidate the interpretation in terms of spin (Fig.~\ref{fig:diags} right). In contrast, the radiation before the decay that is present in other processes does not alter the picture, and can be considered as a quantum map using the formalism of density operators~\cite{Aoude:2025ovu}.

Because quantum tomography relies on a LO interpretation of angular distributions, a possible approach to allow its application to actual measurements is to treat higher-order effects as a {\em background}, to be modeled and subtracted from data alongside other backgrounds contributing to the final state under consideration. Here we will explore this idea for $H \to ZZ \to e^+ e^- \mu^+ \mu^-$, in its simplest implementation. Such type of correction is not strictly necessary for the forthcoming  measurements at the Large Hadron Collider (LHC) using Run $2 + 3$ data, where the statistical uncertainties are still quite large. But it will be compulsory at the high-luminosity upgrade (HL-LHC). 

We remark that considering part of the `signal' differential cross section as a background, and subtracting it from data, is a quite innovative approach for quantum tomography. But it is not the first time such a correction is applied. Previously, the Large Electron Positron (LEP) collaborations provided measurements of $Z$-pole observables $R_b$, $R_c$, etc. after subtracting photon and $\gamma-Z$ interference, among other corrections~\cite{ALEPH:2005ab}. Additional details can be found in Ref.~\cite{ParticleDataGroup:2024cfk}. For single-top $tW$ production several subtraction schemes have been proposed~\cite{Tait:1999cf,Frixione:2008yi} and are applied in measurements in order to remove on-shell $t \bar t$ production from the NLO contribution. In our case, the subtraction term to be applied is obviously gauge-independent because both the LO and NLO differential cross sections are, and likewise their difference $\Delta_\text{NLO} \equiv d\sigma_\text{NLO} - d\sigma_\text{LO}$. 

\section{$H \to ZZ \to e^+ e^- \mu^+ \mu^- (\gamma)$ angular distributions}
\label{sec:2}

In this final state with distinguishable particles the two intermediate $Z$ bosons can be {\em defined} from the two opposite-sign same-flavour pairs, with momenta $p_{Z_1} = p_{\ell_1^+} + p_{\ell_1^-}$, $p_{Z_2} = p_{\ell_2^+} + p_{\ell_2^-}$, $\ell = e,\mu$. For definiteness we label as $Z_1$ the one with largest invariant mass.
We parameterise angular distributions using the helicity basis, a moving reference system with vectors $(\hat r, \hat n, \hat k)$ defined as follows~\cite{Bernreuther:2015yna}:
\begin{itemize}
\item $\hat k$ is chosen in the direction of the $Z_1$ momentum, evaluated either in the Higgs rest frame or the four-lepton c.m. frame (denoted for brevity as $ZZ$ c.m. frame). Both choices are equivalent at LO, but not in the presence of an extra photon. 
\item $\hat r$ is defined as $\hat r = \mathrm{sign}(\cos \theta) (\hat p_p - \cos \theta \hat k)/\sin \theta$, with $\hat p_p = (0,0,1)$ the direction of one proton in the laboratory frame, and $\cos \theta = \hat k \cdot \hat p_p$. The definition for $\hat r$  is the same if we use the direction of the other proton $- \hat p_p$. 
\item $\hat n$ is taken orthogonal, $\hat n = \hat k \times \hat r$.
\end{itemize}
The angular orientation of the decay products can be specified by the polar and azimuthal angles $\Omega_1 = (\theta_1,\phi_1)$ and $\Omega_2 = (\theta_2,\phi_2)$ of the negative lepton momenta in the rest frame of the parent $Z$ boson, measured in the $(\hat r, \hat n, \hat k)$ reference system. The four-dimensional decay angular distribution reads~\cite{Aguilar-Saavedra:2022wam}
\begin{eqnarray}
\frac{1}{\sigma}\frac{d\sigma}{d\Omega_1d\Omega_2} & = & \frac{1}{(4\pi)^2}\left[ 1 +a_{LM}^1 Y_L^M(\Omega_1) + a_{LM}^2  Y_L^M(\Omega_2)   \right. \notag \\
& & \left. + c_{L_1 M_1 L_2 M_2} Y_{L_1}^{M_1}(\Omega_1)Y_{L_2}^{M_2}(\Omega_2)  \right] \,,
\label{ec:dist4D}
\end{eqnarray}
with $Y_L^M$ the usual spherical harmonics, and implicit sum over repeated indices.
At LO, this expansion is general with $L \leq 2$, and the only non-zero coefficients are $a_{20}^{1,2}$, $c_{LML-M}$ with $M \leq L$. At higher orders, other terms and even higher-rank spherical harmonics might in principle have significant contributions. This does not occur in $H \to ZZ \to e^+ e^- \mu^+ \mu^-$ at NLO;\footnote{We have checked it numerically up to $L=4$.} however, if those contributions were present, they could be effectively removed by the subtraction of NLO corrections discussed in the following.

For the calculation of $H \to ZZ \to e^+ e^- \mu^+ \mu^-$ with electroweak corrections we use {\scshape MadGraph5\_aMC@NLO} with the same setup of Ref.~\cite{DelGratta:2025qyp}, namely taking as fundamental parameters for renormalisation
the masses of the weak bosons $M_Z = 91.188$ GeV, $M_W = 80.419$ GeV, and the Fermi constant $G_F = 1.16639 \times 10^{-5}$ GeV$^{-2}$, and using the complex-mass scheme~\cite{Denner:2006ic}.  The masses of the top quark and Higgs boson are taken  as $m_t = 173.3$ GeV, $M_H = 125$ GeV. Photon recombination is performed by clustering photons and leptons into `dressed' leptons if their angular separation $\Delta R \equiv [ (\Delta \phi)^2 + (\Delta \eta)^2]^{1/2}$ is smaller than 0.1. A smaller threshold results in larger differences between NLO and LO predictions for $a$ and $c$ coefficients, while a larger threshold largely recovers the LO values~\cite{DelGratta:2025qyp}.

We present in Table~\ref{tab:ac} the values of $a_{20}^{1,2}$ and $c_{LML-M}$ computed for $H \to ZZ \to e^+ e^- \mu^+ \mu^-$ at LO and NLO. In the latter case, one can choose whether to evaluate $\hat k$ in the Higgs or $ZZ$ c.m. frame. In the latter case, differences with respect to LO are smaller, as pointed out before~\cite{Grossi:2024jae}. One can also choose to veto events for which the photon is sufficiently energetic. The rationale behind this separation is the fact that it actually corresponds to a distinct physical process, which can be experimentally distinguished from $H \to ZZ \to e^+ e^- \mu^+ \mu^-$ provided the photon is sufficiently energetic and separated from the charged leptons. For this, we impose an upper limit of 10 GeV on the photon energy, labelling the results as `exclusive'  in contrast to `inclusive' results where photons of any energy are allowed. Our results for the Higgs rest frame, inclusive in photon momentum, agree very well with those in Ref.~\cite{DelGratta:2025qyp}.

\begin{table}[htb]
\begin{center}
\begin{tabular}{lcccccc}
& LO & \multicolumn{2}{c}{NLO, $H$ frame} & \multicolumn{2}{c}{NLO, $ZZ$ frame} & $\Delta_\text{stat}$ \\
&  & Inclusive & Exclusive & Inclusive & Exclusive  \\
$a_{20}^1$ & -0.663 & -0.571 & -0.649 & -0.632 & -0.658 & 0.075
\\
$a_{20}^2$ & -0.663 & -0.632 &  -0.649 & -0.606 & -0.636 & 0.072
\\
$c_{111-1}$ & 0.296 & 0.046 & 0.057 & 0.048 &  0.057 & 0.22
\\
$c_{1010}$ & -0.181 & -0.013 & -0.015 & -0.010 & -0.015 & 0.25
\\
$c_{222-2}$ & 0.737 & 0.715 & 0.727 & 0.720 & 0.727 & 0.22
\\
$c_{212-1}$ & -1.179 & -1.187 & -1.180 & -1.179 & -1.178 & 0.20
\\
$c_{2020}$ & 1.786 & 1.763 & 1.771 & 1.747 & 1.773 & 0.25
\\
\end{tabular}
\end{center}
\caption{Numerical value of selected coefficients of the distribution (\ref{ec:dist4D}) at LO and NLO, with different assumptions (see the text). The last column is the expected statistical uncertainty at HL-LHC, as obtained in section~\ref{sec:5}.}
\label{tab:ac}
\end{table}

The departures from LO predictions must be contextualised in light of experimental uncertainties at the LHC. While these differences are small compared to the present and near-future precision (see section~\ref{sec:5}), they will reach the $1\sigma$ level for the high luminosity upgrade. For better comparison we include in the last column of Table~\ref{tab:ac} the expected statistical uncertainties for the HL-LHC.

\section{Spin interpretation of angular distributions}
\label{sec:3}

The spin density operator $\rho_{S_1 S_2}$ for two $Z$ bosons\footnote{In $H \to ZZ$ the off-shell propagator includes a scalar degree of freedom; however, when coupled to massless external fermions the scalar component vanishes~\cite{Berge:2015jra} and one can effectively consider $H \to ZZ \to e^+ e^- \mu^+ \mu^-$ as a decay into two spin-1 particles.} can be written as an expansion in irreducible tensors $T^L_M$~\cite{Aguilar-Saavedra:2022wam},
\begin{eqnarray}
\rho_{S_1 S_2} & = & \frac{1}{9}\left[ \mathbb{1}_3 \otimes \mathbb{1}_3 + A^1_{LM} T^L_M \otimes \mathbb{1}_3 + A^2_{LM} \mathbb{1}_3\otimes T^L_M  + C_{L_1 M_1 L_2 M_2}\ T^{L_1}_{M_1} \otimes T^{L_2}_{M_2} \right] \,,
\label{ec:rho}
\end{eqnarray}
with constants $A_{LM}^{1,2}$, $C_{L_1 M_1 L_2 M_2}$. Note that in order for $\rho_{S_1 S_2}$ to be Hermitian, the coefficients must satisfy $A_{LM}^{1,2} = (-1)^M A_{L-M}^{1,2}$, $C_{L_1 M_2 L_2 M_2} = (-1)^{M_1 + M_2} (C_{L1 -M_1 L_2 -M_2})^*$. We normalise $T^L_M$ such that $\Tr\left[T^L_M\; \left(T^L_M\right)^{\dagger}\right] = 3 $, where $\left(T^L_M\right)^{\dagger}=(-1)^M \, T^L_M $. For $L=1$ we have $T^1_{\pm 1}=\mp\sqrt{3}/2 \,  (J_1 \pm i J_2)$ and $T^1_0=\sqrt{3/2} \, J_3$, 
\begin{equation}
T^1_1=\sqrt{\frac{3}{2}}
\left( \! \begin{array}{ccc}
 0 & -1 & 0 \\
 0 & 0 & -1  \\
 0 & 0 & 0
\end{array} \! \right) \,,\quad
T^1_0 = \sqrt{\frac{3}{2}}
\left( \! \begin{array}{ccc}
 1 & 0 & 0 \\
 0 & 0 & 0  \\
 0 & 0 & -1
\end{array} \! \right) \,, \quad
T^1_{-1}=\sqrt{\frac{3}{2}}
\left( \! \begin{array}{ccc}
 0 & 0 & 0 \\
 1 & 0 & 0  \\
 0 & 1 & 0
\end{array} \! \right) \,,
\label{T1}
\end{equation}
where $J_i$ are the usual angular momentum operators.  For $L=2$ they are defined as
\begin{eqnarray}
T^2_{\pm 2} & = & \frac{2}{\sqrt{3}}\, (T_{\pm 1}^1)^2 \,, \nonumber \\
T^2_{\pm 1} & = & \sqrt{\frac{2}{3}} \left[T_{\pm 1}^1T_{0}^1 + T_{0}^1T_{\pm 1}^1\right] \,, \nonumber\\
T^2_0 & = & \frac{\sqrt{2}}{3} \left[T_1^1T_{-1}^1 + T_{-1}^1T_{1}^1+2(T_{0}^1)^2\right] \,.
\end{eqnarray} 
Explicitly,
\begin{align}
& T^2_2 = \sqrt{3}
\left(\!\begin{array}{ccc}
 0 & 0 & 1 \\
 0 & 0 & 0  \\
 0 & 0 & 0
\end{array} \! \right) \,, \quad
T^2_{-2}=\sqrt{3}
\left(\! \begin{array}{ccc}
 0 & 0 & 0 \\
 0 & 0 & 0  \\
 1 & 0 & 0
\end{array} \! \right) \,, \quad
T^2_{1} = \sqrt{\frac{3}{2}}
\left(\! \begin{array}{ccc}
 0 & -1 & 0 \\
 0 & 0 & 1  \\
 0 & 0 & 0
\end{array} \! \right) \,,  \notag \\[2mm]
& T^2_{-1} = \sqrt{\frac{3}{2}}
\left(\! \begin{array}{ccc}
 0 & 0 & 0 \\
 1 & 0 & 0  \\
 0 & -1 & 0
\end{array} \! \right) \,, \quad
T^2_0 = \frac{1}{\sqrt{2}}
\left( \! \begin{array}{ccc}
 1 & 0 & 0 \\
 0 & -2 & 0  \\
 0 & 0 & 1
\end{array} \! \right) \,.
\label{T2}
\end{align}
At LO there is a direct relation between coefficients in the angular distribution (\ref{ec:dist4D}) and coefficients of the density operator (\ref{ec:rho}),
\begin{align}
& a_{LM}^{1,2} = B_L A_{LM}^{1,2}  \,, \notag \\
& c_{L_1 M_1 L_2 M_2} = B_{L_1} B_{L_2} C_{L_1 M_1 L_2 M_2} \,,
\label{ec:acAC}
\end{align}
which is precisely what enables to perform quantum tomography of the $ZZ$ pair. Here, $B_{1,2}$ are constants, $B_1 = - \sqrt{2\pi} \eta_\ell$ and $B_2 = \sqrt{2\pi/5}$, where for leptonic $Z$ decays
\begin{equation}
\eta_\ell = \frac{1-4 s_W^2}{1-4 s_W^2 + 8 s_W^4} \,,
\end{equation}
$s_W$ being the sine of the weak angle~\cite{Aguilar-Saavedra:2017zkn}. As noted in Ref.~\cite{DelGratta:2025qyp}, $\eta_\ell$ is very sensitive to the precise value of $s_W$, because $s_W^2 \simeq 1/4$. For consistency with the parameters used in the NLO calculation we use $s_W^2 = 0.222247$, yielding $\eta_\ell = 0.219$.\footnote{In previous work~\cite{Aguilar-Saavedra:2022wam,Aguilar-Saavedra:2024whi} we have used $\eta_\ell = 0.13$.}
This $\eta_\ell$ factor suppresses the correlation coefficients of spherical harmonics with $L=1$ in the distribution (\ref{ec:dist4D}), namely $c_{1010}$ and $c_{111-1}$. 

Here it is worth pointing out that, already at the LO, a spin interpretation of angular coefficients is not possible in $H \to ZZ \to 4e / 4\mu$~\cite{Aguilar-Saavedra:2024jkj}. In fact, {\em there is no $ZZ$ system} in $4e / 4\mu$ final states due to identical-particle exchange: while in one Feynman diagram an opposite-sign fermion pair is coupled to a $Z$ boson, the same fermion pair couples to different bosons in the other diagram. Albeit to a (numerically) much lesser extent, this also happens in $H \to ZZ \to e^+ e^- \mu^+ \mu^-$ at NLO. If one interprets NLO angular coefficients in Table~\ref{tab:ac} (either column) as spin coefficients using (\ref{ec:acAC}), the resulting density operators are not physical, with three sizeable negative eigenvalues $\lambda \simeq -0.10, -0.19, -0.19$.  Moreover, the necessary and sufficient conditions for spin entanglement~\cite{Aguilar-Saavedra:2022wam}
\begin{equation}
C_{212-1} \neq 0 \quad \text{or} \quad C_{222-2} \neq 0
\label{ec:ent}
\end{equation}
that are specific to a scalar decay into two spin-1 particles, are no longer valid. Ref.~\cite{DelGratta:2025qyp} has addressed this latter issue by the use of other entanglement markers that are not specific for the $0 \to 1 + 1$ spin decay chain. Still, the underlying problem remains: angular correlations {\em cannot} be interpreted as $Z$ boson spin observables when the contribution from diagrams such as those in Fig.~\ref{fig:diags} is non-negligible. Notably, in some diagrams there is not even a $Z$ boson! And this is made manifest by the presence of negative eigenvalues in the density operators. Given the current level of statistical uncertainties (see section~\ref{sec:5}) one may reasonably argue that data is well described by LO and perform quantum tomography as described above, ignoring NLO issues. But this will not be the case at the HL-LHC. Furthermore, independently of experimental uncertainties, it is desirable to have a theoretically sound framework for the interpretation of angular measurements as spin observables. Such a framework can be implemented with the subtraction scheme described in the next section.

\section{Subtraction scheme}
\label{sec:4}

For any `signal' process for which we want to perform quantum tomography, there will necessarily be contributions from backgrounds. These backgrounds have to be subtracted in order to isolate the signal, and their calculation often relies on Monte Carlo simulations, using the Standard Model (SM) prediction. The idea is then to include NLO effects alongside these backgrounds, so that the corrected data can be tested again the LO prediction, for which tomography is well defined. 

At NLO, the differential cross section can be expanded as
\begin{equation}
d\sigma^\text{NLO} = d\Phi_F \;  |\mathcal{M}_F^{(0)}|^2 + d\Phi_F \; 2 \Re \mathcal{M}_F^{(1)} \mathcal{M}_F^{(0)*} + 
\int d\Phi_{F+1} \;  |\mathcal{M}_{F+1}^{(0)}|^2 \,,
\end{equation}
where the first term is the tree-level contribution, the second one the interference between tree-level and virtual corrections, and the third one the real emission diagrams. In the Higgs decay process under consideration, $\mathcal{M}_F^{(0)}$ and $\mathcal{M}_F^{(1)}$ are the amplitudes at tree-level and one loop, and $\mathcal{M}_{F+1}^{(0)}$ is the tree-level amplitude with an extra photon; $\Phi_F$ is the four-lepton phase space, and $\Phi_{F+1}$ the phase space of the four leptons and the photon, which we integrate over the photon degrees of freedom. The difference between NLO and LO differential distributions for the leptons is then
\begin{equation}
\Delta_\text{NLO} \equiv d\Phi_F \; 2 \Re \mathcal{M}_F^{(1)} \mathcal{M}_F^{(0)*} + 
\int d\Phi_{F+1} \;  |\mathcal{M}_{F+1}^{(0)}|^2 \,,
\end{equation}
and is obviously gauge-invariant because both $d\sigma^\text{NLO}$ and $d\sigma^\text{LO}$ are. This correction can be evaluated with Monte Carlo, assuming the SM, and subtracted from (pseudo-)data in order to perform quantum tomography consistently. 

\begin{figure}[t]
\begin{center}
\begin{tabular}{cc}
\includegraphics[height=5.4cm]{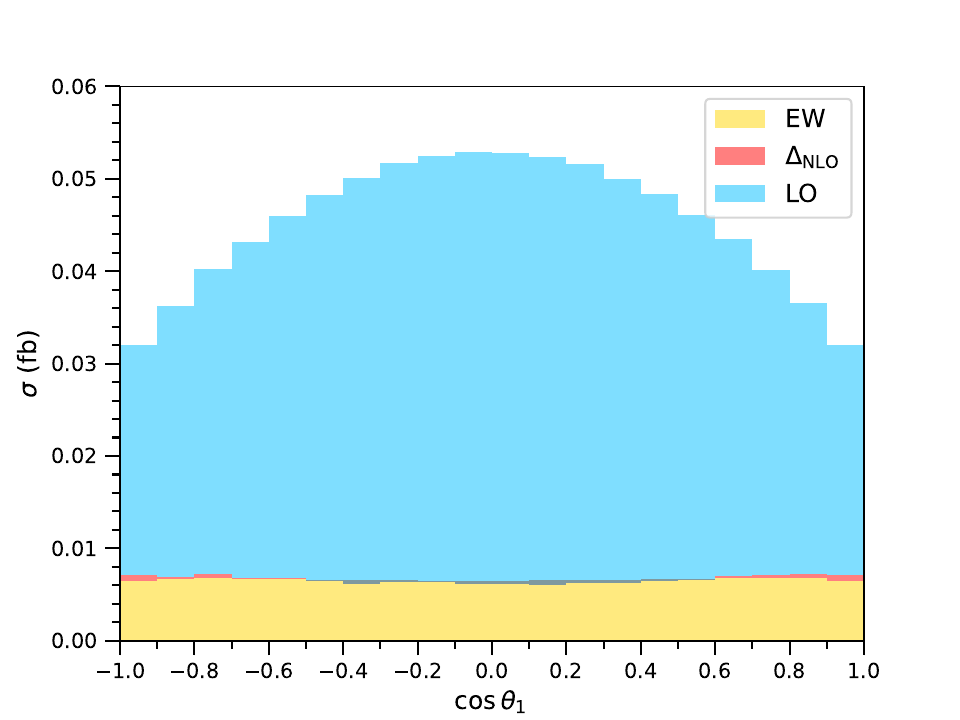}
& \includegraphics[height=5.4cm]{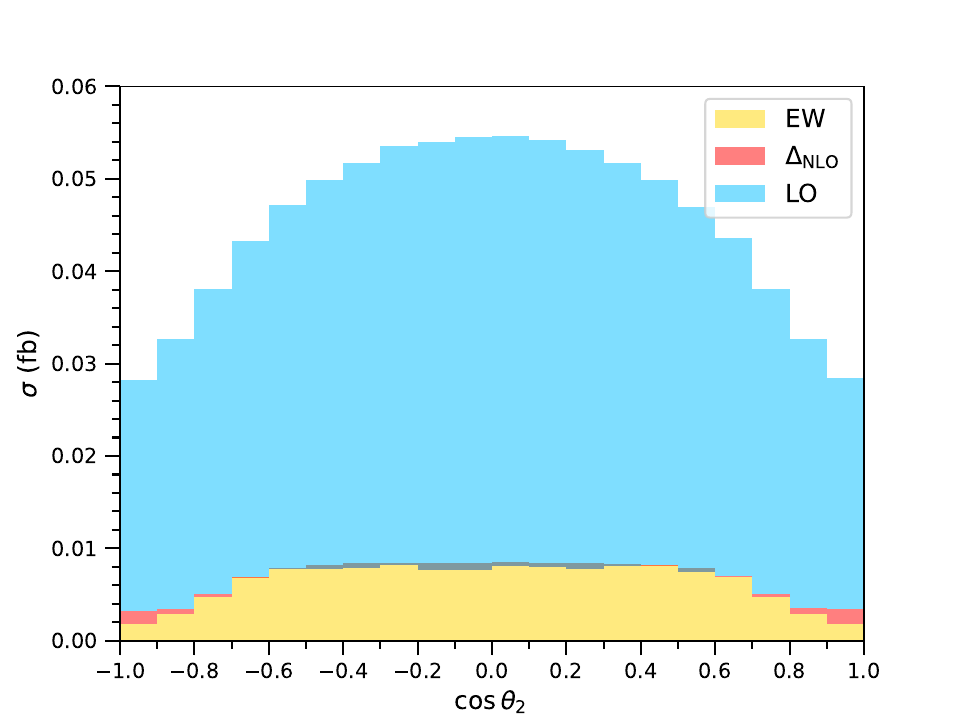} \\
\includegraphics[height=5.4cm]{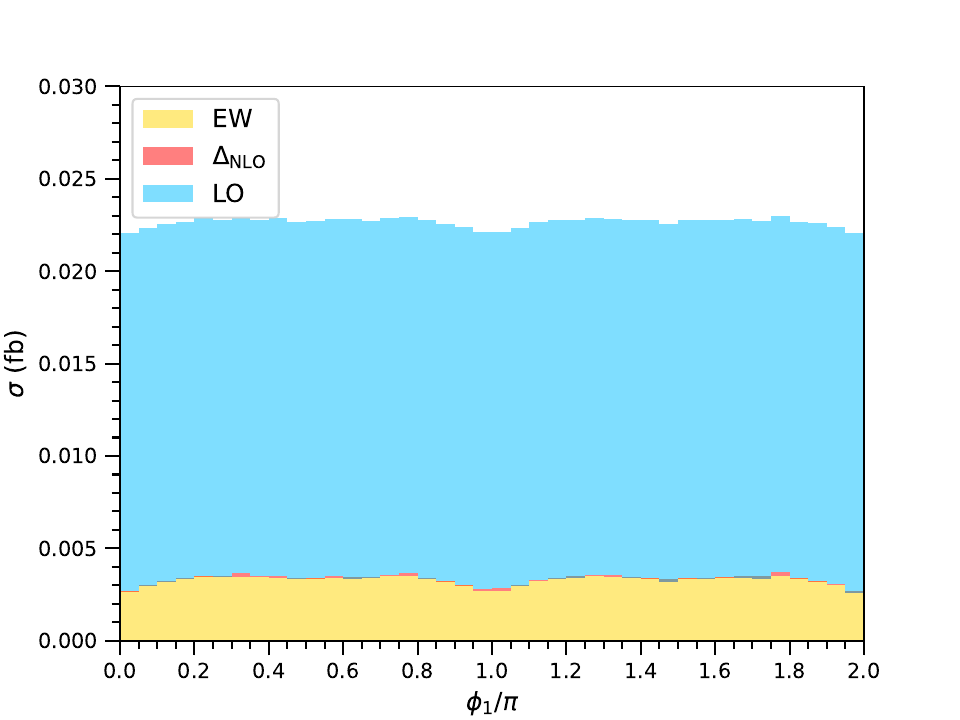}
& \includegraphics[height=5.4cm]{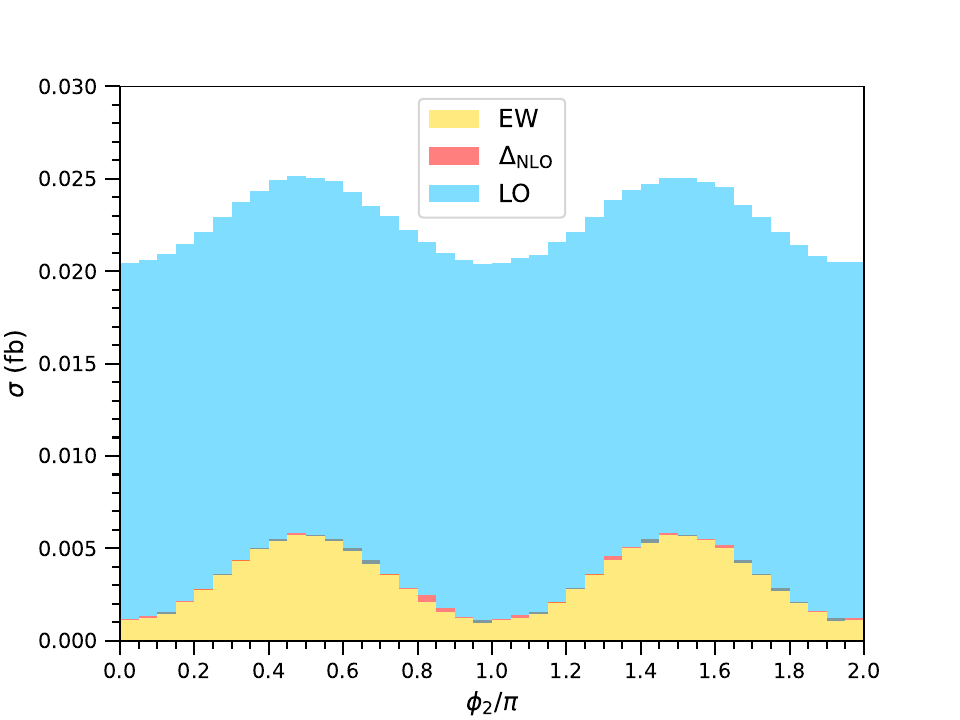} 
\end{tabular}
\caption{Marginalised angular distributions for $\theta_1$, $\theta_2$, $\phi_1$ and $\phi_2$, c.f. Eq.~(\ref{ec:dist4D}), for the electroweak background, $H \to ZZ \to e^+ e^- \mu^+ \mu^+$ at LO, and the $\Delta_\text{NLO}$ difference.}
\label{fig:dist1D}
\end{center}
\end{figure}

The first question that arises is how this correction may affect the uncertainties in the experimental measurement. As we have argued, a background subtraction has nevertheless to be applied. In order to compare the relative size of $\Delta_\text{NLO}$ and the background we have generated with {\scshape MadGraph5\_aMC@NLO} the electroweak four-lepton background $pp \to e^+ e^- \mu^+ \mu^-$, in the invariant mass window $m_{ee\mu\mu} \in [120,130]$ GeV. The tree-level cross section at 14 TeV is 0.50 fb, to which we apply a $K$ factor of 1.1~\cite{Grazzini:2015hta}. On the other hand, the next-to-next-to-leading order Higgs production cross section in gluon gluon fusion is 54.67 pb, with decay branching ratio into $e^+ e^- \mu^+ \mu^-$ of $5.897 \times 10^{-5}$~\cite{Cepeda:2019klc}. We present in Fig.~\ref{fig:dist1D} the  one-dimensional angular distributions corresponding to the four angles $\theta_{1,2}$, $\phi_{1,2}$, for the electroweak background, the LO contribution and the $\Delta_\text{NLO}$ correction. Among the four different possibilities for the NLO calculation shown in Table~\ref{tab:ac}, we choose the Higgs frame, inclusive in photon momenta, for which the deviations are largest.  As it can be seen from Fig.~\ref{fig:dist1D}, the impact of $\Delta_\text{NLO}$ is minimal at least in the marginalised distributions. Note also that $\Delta_\text{NLO}$ can have either sign; for example, for the $\cos \theta_{1,2}$ distributions it is positive near $\theta_{1,2} = 0,\pi$ and negative near $\theta_{1,2} = \pi/2$.

A second point that deserves further discussion is to what extent this correction, evaluated with Monte Carlo assuming the SM, may bias the measurement. In this respect, one has to note that the background is already evaluated for the SM. Furthermore, let us assume the amplitudes receive some correction, either from physics beyond the SM or of other type, 
\begin{align}
& \mathcal{M}_{F}^{(0)} \to \mathcal{\tilde M}_{F}^{(0)} = \mathcal{M}_{F}^{(0)}  + \delta \mathcal{M}_{F}^{(0)}  \,, \notag \\
& \mathcal{M}_{F}^{(1)} \to \mathcal{\tilde M}_{F}^{(1)} =  \mathcal{M}_{F}^{(1)}  + \delta \mathcal{M}_{F}^{(1)}  \,, \notag \\
& \mathcal{M}_{F+1}^{(0)} \to \mathcal{\tilde M}_{F+1}^{(0)} =  \mathcal{M}_{F+1}^{(0)}  + \delta \mathcal{M}_{F+1}^{(0)}  \,.
\end{align}
Provided $\delta \mathcal{M}_{F}^{(1)} \ll  \delta \mathcal{M}_{F}^{(0)}$, $\delta \mathcal{M}_{F+1}^{(0)} \ll  \delta \mathcal{M}_{F}^{(0)}$, $\mathcal{M}_{F}^{(1)}  \ll \mathcal{M}_{F}^{(0)}$, the new differential cross section with the $\Delta_\text{NLO}$ subtraction evaluated for the SM is approximately
\begin{equation}
d\tilde \sigma^\text{NLO} - \Delta_\text{NLO} \simeq d\Phi_F \;  |\mathcal{M}_F^{(0)} + \delta \mathcal{M}_{F}^{(0)} |^2 \,.
\end{equation}
That is, the subtraction of $\Delta_\text{NLO}$ evaluated for the SM keeps the sensitivity to $\delta \mathcal{M}_{F}^{(0)}$.

One can also wonder about the modeling of photon emission, given the fact that the charged leptons as seen by detectors are `dressed leptons' that include collinear photon radiation at all orders, which cannot be resolved. The comparison between theory and data requires a good description of collinear photon radiation. This is so irrespectively of whether theory-data comparison is made at NLO, or NLO corrections are subtracted from data. However, placing a veto on energetic large-angle emission may introduce additional modeling uncertainties. The impact of these uncertainties is process-dependent, and it is expected to be a minor source of uncertainty in processes like $H \to ZZ$ that is largely dominated by statistical uncertainties.

A last point concerns the actual implementation of this subtraction in measurements, because angular distributions in data suffer  from detector effects.  In principle, it is possible to either subtract higher-order effects on unfolded data, or before unfolding. The best choice will depend on the specifics of the analysis, and its investigation is beyond the scope of this work.

\section{Expected experimental uncertainties}
\label{sec:5}

As we have stressed, the differences between LO and NLO predictions for the distribution (\ref{ec:dist4D}) have to be considered in the context of the experimental uncertainties. For this purpose, it is sufficient to estimate the statistical ones, which will likely be dominant for this clean final state. We work at the parton level, but injecting approximate efficiencies of 0.7 for lepton detection, which result in an overall efficiency around $1/4$.  This efficiency accounts for the minimum transverse momentum ($p_T$) thresholds required for lepton detection. We do not include any trigger requirement. The presence of four leptons from the Higgs decay, some of them with significant $p_T$, is expected to fulfill one or many of the trigger conditions for one, two, or three leptons~\cite{trigger}.
We take into account not only the Higgs signal produced in gluon gluon fusion, but also the electroweak four-lepton background, restricted to the Higgs peak $m_{ee\mu\mu} \in [120,130]$ GeV. The background has a moderate effect on the statistical uncertainty because its cross section is around $1/6$ of the signal. (This signal to background ratio is in agreement with the background estimations from recent measurements~\cite{CMS:2023gjz}.)

The statistical uncertainty on the angular coefficients is obtained by performing pseudo-experiments. In each pseudo-experiment, random subsets of $n_S$ signal and $n_B$ background events are drawn from the total event sets, with $n_S,n_B$ the expected numbers for events for each case. The sampling of signal events is tricky, because at the generator level one quarter of the weighted events have negative weights. For a small sample of 100 events (as in Run 2), statistical fluctuations can easily lead to negatively populated bins. We address this difficulty by crafting a $H \to ZZ \to e^+ e^- \mu^+ \mu^-$ unweighted sample with the NLO values of angular observables (c.f. Table~\ref{tab:ac}), using the custom angle rewriting (CAR) method~\cite{Aguilar-Saavedra:2022kgy}. This procedure gives exact results, as long as we do not consider the correlation between the angular observables and other kinematical variables, namely $m_{Z_2}$ and the photon momentum.\footnote{Should one be interested in this correlation, the CAR method could still be applied to subsamples divided in bins, using the values of the $a$ and $c$ coefficients for each bin.} The subtraction terms (four-dimensional distributions) for $\Delta_\text{NLO}$ and the expected EW background are evaluated on large-statistics samples, and applied to the subsets resulting from each pseudo-experiment.

We consider three benchmarks: Run 2 at 13 TeV with a luminosity $L = 139$ fb$^{-1}$; Run $2 + 3$ at 13 / 13.6 TeV with $L = 350$ fb$^{-1}$; and HL-LHC at 14 TeV with $L = 3$ ab$^{-1}$. The cross sections for $gg \to H \to e^+ e^- \mu^+\mu^- (\gamma)$ and the electroweak background are 
\begin{align}
& 13~\text{TeV}: && \sigma_H = 2.86~\text{fb} \,, && \sigma_{EW} = 0.52~\text{fb} \,, \notag \\
& 13.6~\text{TeV}: && \sigma_H = 3.08~\text{fb} \,, && \sigma_{EW} = 0.54~\text{fb} \,, \notag \\
& 14~\text{TeV}: && \sigma_H = 3.22~\text{fb} \,, && \sigma_{EW} = 0.55~\text{fb} \,.
\end{align}
The central values and statistical uncertainties obtained for $N=10000$ pseudo-experiments are collected in Table~\ref{tab:PE}.
As expected, by construction the LO values are recovered within statistical uncertainties. These uncertainties are numerically the same independently of the choice for NLO (Higgs or $ZZ$ c.m., inclusive or exclusive), because they mainly depend on the event sample size --- note that for exclusive measurements with a veto on energetic photons the signal sample size is reduced by only 0.6\%.

\begin{table}[t]
\begin{center}
\begin{tabular}{lcccc}
& Run 2 & Run $2 + 3$ & HL-LHC & LO value \\
$a_{20}^1$ & $-0.66 \pm 0.37$ & $-0.67 \pm 0.23$ & $-0.663 \pm 0.075$ & -0.663 
\\
$a_{20}^2$ & $-0.66 \pm 0.35$ & $-0.66 \pm 0.22$ & $-0.665 \pm 0.072$ & -0.663 
\\
$c_{111-1}$ & $0.30 \pm 1.09$ & $0.29 \pm 0.68$ & $0.30 \pm 0.22$ & 0.296 
\\
$c_{1010}$ & $-0.20 \pm 1.22$ & $-0.21 \pm 0.77$ & $-0.20 \pm 0.25$ & -0.181 
\\
$c_{222-2}$ & $0.73 \pm 1.09$ & $0.75 \pm 0.69$ & $0.74 \pm 0.22$ & 0.737
\\
$c_{212-1}$ & $-1.18 \pm 0.94$ & $-1.18 \pm 0.60$ & $-1.19 \pm 0.20$ & -1.179 
\\
$c_{2020}$ & $1.79 \pm 1.26$ & $1.78 \pm 0.76$ & $1.77 \pm 0.25$ & 1.786 
\end{tabular}
\caption{Central value and statistical uncertainty obtained for the angular observables in Eq.~(\ref{ec:dist4D}) from pseudo-experiments. For comparison, the last column shows the LO value.}
\label{tab:PE}
\end{center}
\end{table}

For Run $2 + 3$ data, the differences between LO and NLO angular coefficients are well below the statistical uncertainty. The largest ones are for  $c_{111-1}$ and $c_{1010}$, up to 0.25 and 0.17, respectively. Given the expected statistical uncertainties of 0.68 and 0.77 in their measurement, those differences amount to $0.36\sigma$ and $0.22\sigma$ respectively.

\section{Summary}
\label{sec:6}

Quantum tomography is well defined only at LO because it entails identifying the `mother' particles (top quarks, $W/Z$ bosons, etc.) that produce the ones seen in the detector (electrons, muons, jets, etc.) Beyond LO this association is ill-defined; therefore, in processes where higher-order effects are significant, a consistent framework is necessary for this interpretation. In this paper we have proposed considering higher-order corrections as a background to be subtracted from data. Doing so, the interpretation of angular observables in terms of spin and spin correlations of intermediate particles is legitimate. Or course, should a significant discrepancy between data and predictions be observed at any level, its possible source (mismodeling, new physics, etc.) must be be investigated and the consistency of an interpretation in terms of spin checked.

We have illustrated this procedure for $H \to ZZ \to e^+ e^- \mu^+ \mu^-$, which is of high interest in view of upcoming measurements of qutrit entanglement. Although the necessary corrections are small when compared to the expected statistical uncertainty, they will be important for future measurements at the HL-LHC.

\section*{Acknowledgements}

I thank M. del Gratta for help with Madgraph, P.P. Giardino and K. Asteriadis for useful discussions, and the CERN Theory Department for hospitality during the realisation of this work. This work has been supported by the Spanish Research Agency (Agencia Estatal de Investigaci\'on) through projects PID2022-142545NB-C21,  and CEX2020-001007-S funded by MCIN/AEI/10.13039/501100011033.


\begin{thebibliography}{99}



\bibitem{Afik:2020onf}
Y.~Afik and J.~R.~M.~de Nova,
Eur. Phys. J. Plus \textbf{136} (2021) no.9, 907
[arXiv:2003.02280 [quant-ph]].

\bibitem{Fabbrichesi:2021npl}
M.~Fabbrichesi, R.~Floreanini and G.~Panizzo,
Phys. Rev. Lett. \textbf{127} (2021) no.16, 16
[arXiv:2102.11883 [hep-ph]].

\bibitem{Barr:2021zcp}
A.~J.~Barr,
Phys. Lett. B \textbf{825} (2022), 136866
[arXiv:2106.01377 [hep-ph]].

\bibitem{Severi:2021cnj}
C.~Severi, C.~D.~Boschi, F.~Maltoni and M.~Sioli,
Eur. Phys. J. C \textbf{82} (2022) no.4, 285
[arXiv:2110.10112 [hep-ph]].

\bibitem{Afik:2022kwm}
Y.~Afik and J.~R.~M.~de Nova,
Quantum \textbf{6} (2022), 820
[arXiv:2203.05582 [quant-ph]].

\bibitem{Aguilar-Saavedra:2022uye}
J.~A.~Aguilar-Saavedra and J.~A.~Casas,
Eur. Phys. J. C \textbf{82} (2022) no.8, 666
[arXiv:2205.00542 [hep-ph]].

\bibitem{Afik:2022dgh}
Y.~Afik and J.~R.~M.~de Nova,
Phys. Rev. Lett. \textbf{130} (2023) no.22, 221801
[arXiv:2209.03969 [quant-ph]].

\bibitem{Aguilar-Saavedra:2022wam}
J.~A.~Aguilar-Saavedra, A.~Bernal, J.~A.~Casas and J.~M.~Moreno,
Phys. Rev. D \textbf{107} (2023) no.1, 016012
[arXiv:2209.13441 [hep-ph]].

\bibitem{Aguilar-Saavedra:2022mpg}
J.~A.~Aguilar-Saavedra,
Phys. Rev. D \textbf{107} (2023) no.7, 076016
[arXiv:2209.14033 [hep-ph]].


\bibitem{Altakach:2022ywa}
M.~M.~Altakach, P.~Lamba, F.~Maltoni, K.~Mawatari and K.~Sakurai,
Phys. Rev. D \textbf{107} (2023) no.9, 093002
[arXiv:2211.10513 [hep-ph]].

\bibitem{Fabbrichesi:2023cev}
M.~Fabbrichesi, R.~Floreanini, E.~Gabrielli and L.~Marzola,
Eur. Phys. J. C \textbf{83} (2023) no.9, 823
[arXiv:2302.00683 [hep-ph]].


\bibitem{Dong:2023xiw}
Z.~Dong, D.~Gon\c{c}alves, K.~Kong and A.~Navarro,
Phys. Rev. D \textbf{109} (2024) no.11, 115023
[arXiv:2305.07075 [hep-ph]].

\bibitem{Morales:2023gow}
R.~A.~Morales,
Eur. Phys. J. Plus \textbf{138} (2023) no.12, 1157
[arXiv:2306.17247 [hep-ph]].

\bibitem{Aguilar-Saavedra:2023hss}
J.~A.~Aguilar-Saavedra,
Phys. Rev. D \textbf{108} (2023) no.7, 076025
[arXiv:2307.06991 [hep-ph]].

\bibitem{Fabbri:2023ncz}
F.~Fabbri, J.~Howarth and T.~Maurin,
Eur. Phys. J. C \textbf{84} (2024) no.1, 20
[arXiv:2307.13783 [hep-ph]].


\bibitem{Aguilar-Saavedra:2023lwb}
J.~A.~Aguilar-Saavedra,
Phys. Lett. B \textbf{848} (2024), 138409
[arXiv:2308.07412 [hep-ph]].

\bibitem{Han:2023fci}
T.~Han, M.~Low and T.~A.~Wu,
JHEP \textbf{07} (2024), 192
[arXiv:2310.17696 [hep-ph]].

\bibitem{Ehataht:2023zzt}
K.~Ehat\"aht, M.~Fabbrichesi, L.~Marzola and C.~Veelken,
Phys. Rev. D \textbf{109} (2024) no.3, 032005
[arXiv:2311.17555 [hep-ph]].

\bibitem{Aguilar-Saavedra:2024fig}
J.~A.~Aguilar-Saavedra and J.~A.~Casas,
Phys. Rev. Lett. \textbf{133} (2024) no.11, 111801
[arXiv:2401.06854 [hep-ph]].

\bibitem{Aguilar-Saavedra:2024hwd}
J.~A.~Aguilar-Saavedra,
Phys. Rev. D \textbf{109} (2024) no.9, 096027
[arXiv:2401.10988 [hep-ph]].

\bibitem{Aguilar-Saavedra:2024vpd}
J.~A.~Aguilar-Saavedra,
Phys. Lett. B \textbf{855} (2024), 138849
[arXiv:2402.14725 [hep-ph]].

\bibitem{Aguilar-Saavedra:2024whi}
J.~A.~Aguilar-Saavedra,
Phys. Rev. D \textbf{109} (2024) no.11, 113004
[arXiv:2403.13942 [hep-ph]].

\bibitem{Maltoni:2024csn}
F.~Maltoni, C.~Severi, S.~Tentori and E.~Vryonidou,
JHEP \textbf{09} (2024), 001
[arXiv:2404.08049 [hep-ph]].

\bibitem{Fabbrichesi:2024wcd}
M.~Fabbrichesi and L.~Marzola,
Phys. Rev. D \textbf{110} (2024) no.7, 076004
[arXiv:2405.09201 [hep-ph]].

\bibitem{Altomonte:2024upf}
C.~Altomonte, A.~J.~Barr, M.~Eckstein, P.~Horodecki and K.~Sakurai,
[arXiv:2412.01892 [hep-ph]].

\bibitem{Han:2024ugl}
T.~Han, M.~Low, N.~McGinnis and S.~Su,
[arXiv:2412.21158 [hep-ph]].

\bibitem{Han:2025ewp}
T.~Han, M.~Low and Y.~Su,
[arXiv:2501.04801 [hep-ph]].

\bibitem{Bernal:2025zqq}
A.~Bernal, J.~A.~Casas and J.~Falceto,
[arXiv:2503.17297 [quant-ph]].

\bibitem{Zhang:2025mmm}
Y.~Zhang, B.~H.~Zhou, Q.~B.~Liu, S.~Li, S.~C.~Hsu, T.~Han, M.~Low and T.~A.~Wu,
[arXiv:2504.01496 [hep-ph]].




\bibitem{Fabbrichesi:2025ifv}
M.~Fabbrichesi, R.~Floreanini, E.~Gabrielli and L.~Marzola,
[arXiv:2503.14587 [hep-ph]].

\bibitem{Fabbrichesi:2025rsg}
M.~Fabbrichesi, R.~Floreanini, E.~Gabrielli and L.~Marzola,
[arXiv:2504.12382 [hep-ph]].



\bibitem{Aguilar-Saavedra:2024jkj}
J.~A.~Aguilar-Saavedra,
[arXiv:2411.13464 [hep-ph]].




\bibitem{Kane:1991bg}
G.~L.~Kane, G.~A.~Ladinsky and C.~P.~Yuan,
Phys. Rev. D \textbf{45} (1992), 124-141

\bibitem{Aguilar-Saavedra:2015yza}
J.~A.~Aguilar-Saavedra and J.~Bernabeu,
Phys. Rev. D \textbf{93} (2016) no.1, 011301
[arXiv:1508.04592 [hep-ph]].

\bibitem{Aguilar-Saavedra:2017zkn}
J.~A.~Aguilar-Saavedra, J.~Bernab\'eu, V.~A.~Mitsou and A.~Segarra,
Eur. Phys. J. C \textbf{77} (2017) no.4, 234
[arXiv:1701.03115 [hep-ph]].

\bibitem{Rahaman:2021fcz}
R.~Rahaman and R.~K.~Singh,
Nucl. Phys. B \textbf{984} (2022), 115984
[arXiv:2109.09345 [hep-ph]].

\bibitem{Ashby-Pickering:2022umy}
R.~Ashby-Pickering, A.~J.~Barr and A.~Wierzchucka,
JHEP \textbf{05} (2023), 020
[arXiv:2209.13990 [quant-ph]].

\bibitem{Bernal:2023jba}
A.~Bernal,
Phys. Rev. D \textbf{109} (2024) no.11, 116007
[arXiv:2310.10838 [hep-ph]].


\bibitem{ATLAS:2023fsd}
G.~Aad \textit{et al.} [ATLAS],
Nature \textbf{633} (2024) no.8030, 542-547
[arXiv:2311.07288 [hep-ex]].

\bibitem{CMS:2024pts}
A.~Hayrapetyan \textit{et al.} [CMS],
Rept. Prog. Phys. \textbf{87} (2024) no.11, 117801
[arXiv:2406.03976 [hep-ex]].

\bibitem{CMS:2024zkc}
A.~Hayrapetyan \textit{et al.} [CMS],
Phys. Rev. D \textbf{110} (2024) no.11, 112016
[arXiv:2409.11067 [hep-ex]].





\bibitem{Grossi:2024jae}
M.~Grossi, G.~Pelliccioli and A.~Vicini,
JHEP \textbf{12} (2024), 120
[arXiv:2409.16731 [hep-ph]].

\bibitem{DelGratta:2025qyp}
M.~Del Gratta, F.~Fabbri, P.~Lamba, F.~Maltoni and D.~Pagani,
[arXiv:2504.03841 [hep-ph]].


\bibitem{Aoude:2025ovu}
R.~Aoude, A.~J.~Barr, F.~Maltoni and L.~Satrioni,
[arXiv:2504.07030 [quant-ph]].




\bibitem{ALEPH:2005ab}
S.~Schael \textit{et al.} [ALEPH, DELPHI, L3, OPAL, SLD, LEP Electroweak Working Group, SLD Electroweak Group and SLD Heavy Flavour Group],
Phys. Rept. \textbf{427}, 257-454 (2006)
[arXiv:hep-ex/0509008 [hep-ex]].

\bibitem{ParticleDataGroup:2024cfk}
S.~Navas \textit{et al.} [Particle Data Group],
``Review of particle physics,''
Phys. Rev. D \textbf{110}, no.3, 030001 (2024)

\bibitem{Tait:1999cf}
T.~M.~P.~Tait,
Phys. Rev. D \textbf{61} (1999), 034001
[arXiv:hep-ph/9909352 [hep-ph]].

\bibitem{Frixione:2008yi}
S.~Frixione, E.~Laenen, P.~Motylinski, B.~R.~Webber and C.~D.~White,
JHEP \textbf{07} (2008), 029
[arXiv:0805.3067 [hep-ph]].




\bibitem{Bernreuther:2015yna}
W.~Bernreuther, D.~Heisler and Z.~G.~Si,
JHEP \textbf{12} (2015), 026
[arXiv:1508.05271 [hep-ph]].

\bibitem{Alwall:2014hca}
J.~Alwall, R.~Frederix, S.~Frixione, V.~Hirschi, F.~Maltoni, O.~Mattelaer, H.~S.~Shao, T.~Stelzer, P.~Torrielli and M.~Zaro,
JHEP \textbf{07} (2014), 079
[arXiv:1405.0301 [hep-ph]].

\bibitem{Denner:2006ic}
A.~Denner and S.~Dittmaier,
Nucl. Phys. B Proc. Suppl. \textbf{160} (2006), 22-26
[arXiv:hep-ph/0605312 [hep-ph]].

\bibitem{Berge:2015jra}
S.~Berge, S.~Groote, J.~G.~K\"orner and L.~Kaldam\"ae,
Phys. Rev. D \textbf{92} (2015) no.3, 033001
[arXiv:1505.06568 [hep-ph]].

\bibitem{Grazzini:2015hta}
M.~Grazzini, S.~Kallweit and D.~Rathlev,
Phys. Lett. B \textbf{750} (2015), 407-410
[arXiv:1507.06257 [hep-ph]].

\bibitem{Cepeda:2019klc}
M.~Cepeda, S.~Gori, P.~Ilten, M.~Kado, F.~Riva, R.~Abdul Khalek, A.~Aboubrahim, J.~Alimena, S.~Alioli and A.~Alves, \textit{et al.}
CERN Yellow Rep. Monogr. \textbf{7} (2019), 221-584
[arXiv:1902.00134 [hep-ph]].

\bibitem{trigger}
ATLAS Collaboration, Trigger menu in 2017, Technical Report No. ATL-DAQ-PUB-2018-002, CERN, Geneva, 2018.

\bibitem{CMS:2023gjz}
A.~Hayrapetyan \textit{et al.} [CMS],
JHEP \textbf{08} (2023), 040
[arXiv:2305.07532 [hep-ex]].

\bibitem{Aguilar-Saavedra:2022kgy}
J.~A.~Aguilar-Saavedra,
Phys. Rev. D \textbf{106} (2022) no.11, 115021
[arXiv:2208.00424 [hep-ph]].



\end{thebibliography}
\end{document}